\documentclass[
    ,final            
  ]
  {aipproc}

\layoutstyle{6x9}
\usepackage{graphicx}
\usepackage{amsmath}
\usepackage{amsfonts}

\newcommand{\beq}{\begin{equation}}
\newcommand{\eeq}{\end{equation}}
\def\lesssim{\mathrel{\hbox{\rlap{\hbox{\lower4pt\hbox{$\sim$}}}\hbox{$<$}}}}
\def\gtrsim{\mathrel{\hbox{\rlap{\hbox{\lower4pt\hbox{$\sim$}}}\hbox{$>$}}}}

\begin{document}

\title{Gravitational Waves from Compact Objects Accreting onto Active
  Galactic Nuclei}


\author{Jeremy Schnittman}{
  address={Physics Department, University of Maryland, College Park,
  MD 20742}
}

\author{G\"unter Sigl}{
  address={AstroParticules et Cosmologie,
    11 place Marcelin Berthelot, F-75005 Paris, France\\
    Institut d'Astrophysique de Paris, 98bis Boulevard Arago,
    75014 Paris, France}
}

\author{Alessandra Buonanno}{
  address={Physics Department, University of Maryland, College Park,
  MD 20742}
}

\begin{abstract}
We consider a model in which massive stars form in a self-gravitating
accretion disk around an active galactic nucleus. These stars
may evolve and collapse to form compact objects on a time scale
shorter than the accretion time, thus producing an important family of
sources for LISA. Assuming the compact object formation/inspiral rate
is proportional to the steady-state gas accretion rate, we use the
observed extra-galactic X-ray luminosity function to estimate expected
event rates and signal strengths. We find that these sources will
produce a continuous low-frequency background detectable by LISA if
more than $\gtrsim 1\%$ of the accreted matter is in the form of compact
objects. For compact
objects with $m \gtrsim 10 M_\odot$, the last stages of the inspiral
events should be resolvable above a few mHz, at a rate of $\sim 10-100$
per year.
\end{abstract}

\maketitle


\section{INTRODUCTION}

One of the major challenges that will face gravitational wave (GW)
astronomers is the successful discrimination
between instrumental noise, stochastic GW backgrounds,
and individual, resolvable GW sources. For example, the population of
galactic white dwarf binaries in close orbits will provide a major
contribution to the LISA noise curve in the $0.1-1$ mHz band
\cite{BH}. At the same time, this confusion ``noise'' can also be
treated as a signal, and its shape and amplitude will provide
important information about the distribution and properties of white
dwarf binaries in the galaxy.

It has recently been proposed that, in the self-gravitating accretion
disks of active galactic nuclei (AGN), massive stars could form and
evolve, eventually collapsing into compact objects and merging with
the central black hole \cite{Goodman:2003sf,Levin}.
At different stages in the inspiral evolution, this population will
contribute to the GW background confusion or alternatively
produce individual, resolvable chirp signals. It is therefore a matter
of theoretical and practical interest to understand the nature
of such a population.

In this paper we derive a relationship between the
observable electro-magnetic (EM) emission and the predicted GW
emission from AGN, following the procedures and summarizing the
results outlined in our longer paper \cite{sigl}. In particular, we use
the hard X-ray luminosity function of Ueda et al.~\cite{uaom} to infer
the accretion history of supermassive black holes (SMBHs) out to
redshifts of $z\sim 3$. Then we assume a few simple scaling factors,
such as the average (Eddington-scaled) accretion rate and the
efficiency of converting accretion energy to X-rays, and derive the
GW spectrum that might be seen by LISA. 

Depending on the specific model parameters, we find this background
could be an important class of LISA sources, similar in strength and
event rates to extreme mass-ratio
inspirals from captured compact objects \cite{barrack}. As in
those sources,
here too it is a matter of preference as to whether the background
should be thought of as signal or noise. But for higher masses
(perhaps as large as $m\simeq 10^5 M_\odot$ \cite{Goodman:2003sf}),
disk-embedded compact objects should produce individual, resolvable
inspiral events with high signal-to-noise over a wide band of
frequencies. 

\section{The hard X-ray luminosity function of AGN}\label{Xray_lf}
We begin with a short discussion of notation. The results
derived below include a number of dimensionless parameters, most of
which can take values between 0 and 1. We divide these parameters into
two general classes: efficiencies and fractions.
Efficiencies, denoted by $\eta$, are believed to be
determined by more fundamental physics, and typically have more stringent
lower- and upper-limits. Fractions, denoted by $f$, are more
model-dependent parameters and less-well known than the efficiency
parameters, and thus have a larger range of acceptable values. 
A summary of these model
parameters appears in Table \ref{glossary}.

\begin{table}
\caption{\label{glossary} Glossary of dimensionless parameters, with
allowable and preferred values}
\begin{tabular}{ccccp{8cm}}
\hline
\hline
symbol & min & max & preferred & description \\
\hline 
$f_{\rm acc}$     & 0   & 1   & 1   & fraction of SMBH mass due to
accreted gas \\
$f_{\rm co}$      & 0   & 1   & 0.01 & fraction of
SMBH mass due to accreted compact objects \\
$f_{\rm X}$             & 0   & 1   & 0.03 & fraction of EM radiation in
X-rays \\
$f_{\rm Edd}$     & 0   & $\gtrsim 1$ & 0.1 & typical
fraction of Eddington luminosity/accretion rate \\
\hline 
$\eta_{\rm em}$   & 0   & 1   & 0.2 & efficiency of
converting accreting gas to EM radiation \\
$\eta_{\rm gw}$   & 0   & 1 & 0.2 & efficiency of
converting compact objects to GW radiation \\
\end{tabular}
\end{table}

A growing consensus has been forming that SMBHs grow almost
exclusively by accretion, suggesting $f_{\rm acc}\simeq1$ (see, e.g.\
\cite{yu}). A corollary
of this assumption is that most AGN should be rapidly spinning, with
dimensionless spin parameters of $a/M \simeq 0.9-0.998$, giving
EM efficiencies for a radiative disk of $\eta_{\rm em} \simeq
0.15-0.32$ \cite{thorne}. Similarly,
assuming the compact objects are on
circular, adiabatic orbits, we set $\eta_{\rm gw}=\eta_{\rm em}$,
ignoring any EM or GW emission from the plunging region.

We write the X-ray luminosity
$L_{\rm X}$ as a fraction $f_{\rm X}$
of the bolometric luminosity, which in turn is a fraction $f_{\rm
  Edd}$ of the Eddington luminosity $L_{\rm Edd}$:
\begin{equation}\label{Lx}
  L_{\rm X}=f_{\rm X} f_{\rm Edd} L_{\rm Edd}(M) = f_{\rm X}
  \eta_{\rm em}\dot{M}_{\rm acc}c^2. 
\end{equation}
The Eddington limit is a function only of the SMBH mass: $L_{\rm
Edd}(M)=1.3\times10^{38}(M/M_\odot)$. Over the range of redshifts
and luminosities we are probing, typical accretion rates are estimated
to be $f_{\rm Edd}\sim0.1$, but could conceivably be even greater than
unity \cite{marconi}. Since the energy density in the cosmic
infrared background (also dominated by AGN) is about 30 times greater
than the X-ray background, we set $f_{\rm X} = 0.03$ \cite{ressell}.
Lastly, the parameter $f_{\rm co}$ is the fraction of total mass
accreted in the form of compact objects. Since the astrophysical
mechanisms that actually determine this fraction are not yet well understood,
we set it to a conservative value of $0.01$. If it were much higher,
the disk would be entirely fragmented and thus not efficiently emit EM
radiation. And as we will see below, a value much below $0.01$ would result
in a GW signal undetectable by LISA.

We approximate the intrinsic (i.e.\ directly produced by the accretion
process, and before reprocessing and/or partial absorption within the
host galaxy) X-ray
luminosity function per comoving volume after Ref.~\cite{uaom},
roughly giving a broken power-law distribution with higher average
luminosities at higher redshift. The fiducial
values for the model parameters and the observed range of luminosities
correspond to SMBH masses in the range $10^6M_\odot\lesssim
M\lesssim10^{10}M_\odot$, consistent with the masses inferred from
observations of velocity dispersions. 

\section{The gravitational wave spectrum}\label{GW_signal}

We begin by considering the inspiral of a single compact object of
mass $m$ onto a SMBH of mass $M\gg m$ and specific angular momentum
$a$. Using geometrized units such that
$G=c=1$, a particle on a circular, equatorial orbit around a Kerr
black hole has an orbital frequency (as measured by an observer at
infinity) of
\begin{equation}
f_{\rm orb}(r) = \frac{\sqrt{M}}{2\pi(r^{3/2}+ a\sqrt{M})}
\end{equation}
and specific energy
\begin{equation}\label{E_r}
\frac{E(r)}{m} = \frac{r^2-2Mr+ a\sqrt{Mr}}{r(r^2-3Mr+
2a\sqrt{Mr})^{1/2}}.
\end{equation}
Thus the total energy emitted in gravitational waves down to a radius
$r$ is $E_{\rm gw}(r)=m-E(r)$.
The GW energy emitted between frequency $f$ and $f + df$ for such
an event is 
\begin{equation}\label{dEdf}
\frac{dE_{\rm gw}}{df} = \frac{dE_{\rm gw}}{dr}
\left(\frac{df}{dr}\right)^{-1}.
\end{equation}
Here, we restrict the GW emission to the quadrupole formula for
circular geodesic orbits, thus 
we consider only GW frequencies at twice the orbital frequencies
$(f=2f_{\rm orb})$. 

\begin{figure}
\caption{\label{single_inspiral} Characteristic GW strain amplitudes
  for individual inspiral
  events, where a black hole with $m=10M_\odot$ merges with a SMBH of mass
  $M=10^6, 10^7, 10^8 M_\odot$ at a redshift of $z=1$. For each value
  of $M$, we show the spectra for two spin values, $a/M=0$ ({\it
  solid}) and $0.95$ ({\it dashed}). The {\it dot-dashed} and {\it
  dotted} lines are the sky-averaged LISA noise curves with and
  without the contributions from galactic binaries, respectively.}
\scalebox{0.6}{\includegraphics{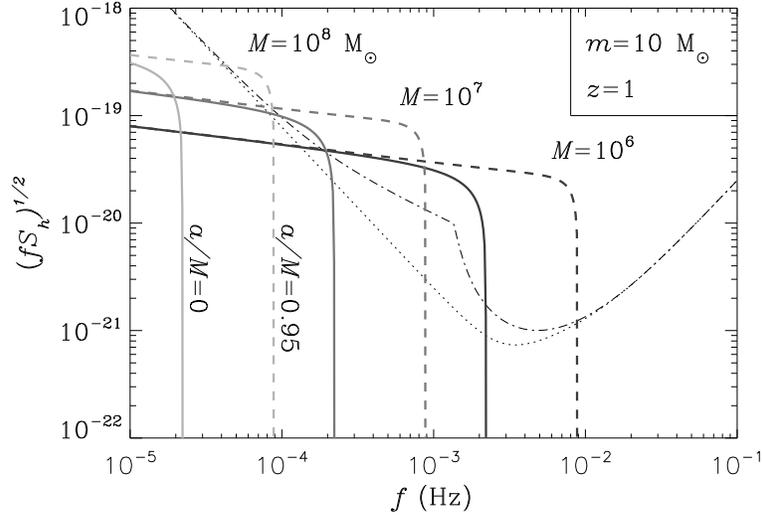}}
\end{figure}

We will generally want to restrict equation (\ref{dEdf}) to a range of
frequencies $f_{\rm min} \le f \le f_{\rm max}$, where $f_{\rm min}$
is determined by the LISA sensitivity and $f_{\rm max}$ is the GW
frequency at the inner-most stable circular orbit (ISCO). The ISCO
frequency in turn is determined solely by the SMBH mass and spin,
giving $f_{\rm max} \simeq 4-30$ mHz for $M=10^6
M_\odot$. In Figure \ref{single_inspiral} we show the characteristic
strain spectrum for
a single inspiral event for a range of black hole masses and spins.

Integrated over redshift $z$, the observed GW energy density per
logarithmic frequency is given by (e.g.\ \cite{phinney})
\begin{equation}\label{Omega1}
  \frac{d\rho_{\rm gw}(f)}{d \ln f} 
  =\int_0^\infty dz\,
  \frac{R(z)}{1+z}\left|\frac{dt}{dz}\right|
  \frac{dE_{\rm gw}}{d\ln f_z}(f_z),
\end{equation}
where $f_z\equiv(1+z)f$, and $R(z)$ is the rate of inspiral
events per comoving volume. Cosmology enters with the term
$|dt/dz|=[(1+z)H(z)]^{-1}$, where for a flat geometry,
\begin{equation}\label{cosmo}
  H(z)= H_0
  \left[\Omega_{\rm M}(1+z)^3+\Omega_{\Lambda}\right]^{1/2}.
\end{equation}
Throughout this paper we will assume a standard $\Lambda$CDM universe
with $\Omega_{\rm M}=0.3$, $\Omega_{\Lambda}=0.7$, and $H_0=72~{\rm
  km}~{\rm s}^{-1}~{\rm Mpc}^{-1}$.

The event rate for a single AGN is simply the X-ray luminosity divided
by the total X-ray energy emitted between inspiral events: $E_{\rm X} =
E_{\rm gw}f_{\rm X}/f_{\rm co}$. Integrating over the luminosity
distribution function, we get
\begin{equation}\label{R_z}
  R(z) = \frac{f_{\rm co}}{f_{\rm X}}\int dL_{\rm X} \frac{dn(L_{\rm X},z)}{d\ln L_{\rm X}}
  \frac{1}{E_{\rm gw}},
\end{equation}
where $dn(L_{\rm X},z)/d \ln L_{\rm X}$ is the intrinsic luminosity
distribution function in units of Mpc$^{-3}$.
Combining equations (\ref{Omega1}) and (\ref{R_z}), the total
(time-averaged) gravitational wave spectrum is 
\begin{equation}\label{Omega3}
  \frac{d \rho_{\rm gw}(f)}{d \ln f}= \frac{f_{\rm co}}{f_{\rm X}}
  \int_0^\infty \left|\frac{dt}{dz}\right|\,\frac{dz}{1+z}
  \int_{L_{\rm min}}^{L_{\rm max}} dL_{\rm X}
  \frac{dn}{d \ln L_{\rm X}}
  \frac{1}{E_{\rm gw}} \frac{dE_{\rm gw}}{d\ln f_z}(f_z).
\end{equation}
The GW spectrum $E_{\rm gw}(f)$ from each individual AGN is a function
of the SMBH mass, which in turn is determined by the X-ray luminosity
through equation (\ref{Lx}). Note that the integrated spectrum is
independent of $m$, as long as $m$ is small
enough so that the inspiral waveform cannot be individually
resolved. One measure of this resolvability is the {\it duty cycle},
described in the next section.

Following Refs.~\cite{barrack,finn}, we will want to compare directly
the background defined in equation (\ref{Omega3}) to the
spectral density of the detector noise $S_{\rm n}(f)$, which has units of
inverse frequency. In this case, $\sqrt{f S_{\rm n}(f)}$ will be a
dimensionless strain. Averaging over the entire sky, weighted by the
LISA antenna pattern, gives
\begin{equation}
S_h(f) = \frac{4}{\pi}\,\frac{1}{f^{3}}\,
\frac{d \rho_{\rm gw}(f)}{d \ln f}.\label{Srho}
\end{equation}
Throughout the paper we use the so-called sky and detector averaged
instrumental spectral density for LISA, augmented by the white-dwarf
galactic confusion noise, as given in \cite{barrack}.

\begin{figure}
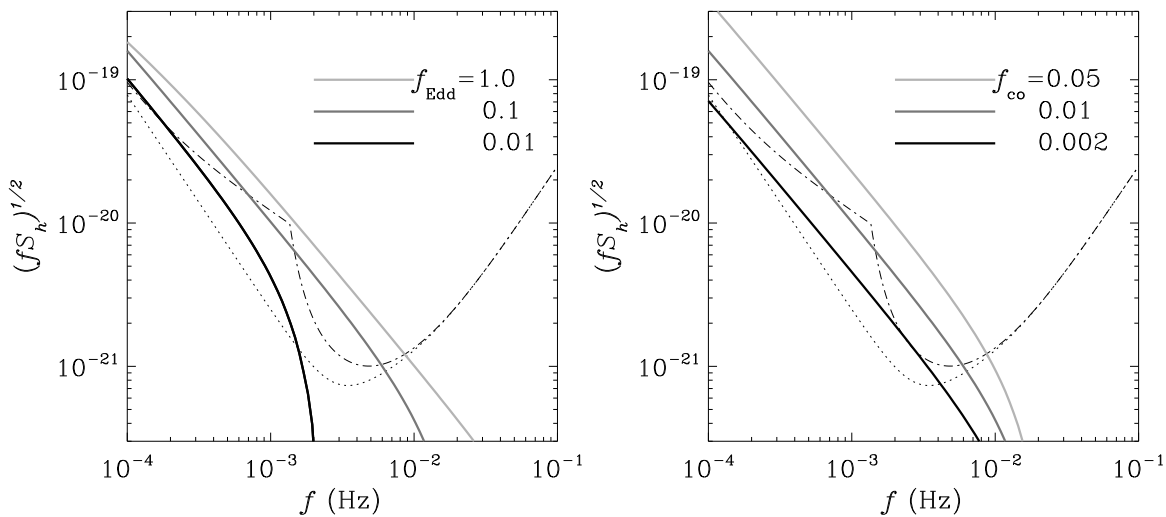

\caption{\label{fiducial} Dependence of the GW background on the
  model parameters $f_{\rm Edd}$ ($left$) and $f_{\rm co}$ ($right$). The
  fiducial values for the model give a background signal comparable to
  the LISA noise curve, including unresolved galactic white dwarf
  binaries.} 
\scalebox{0.6}{\includegraphics{fig2a.epsi}}
\scalebox{0.6}{\includegraphics{fig2b.epsi}}
\end{figure} 

In Figure \ref{fiducial} we plot the nominal GW background from
equation (\ref{Omega3}) with the model parameters listed in Table
1. Also shown are the effects of varying the accretion rate $f_{\rm
  Edd}$ (Fig.\ \ref{fiducial}a) and the fraction of accreted mass in
compact objects $f_{\rm co}$ (Fig.\
\ref{fiducial}b). By increasing $f_{\rm Edd}$, the effect is to
reduce the AGN mass inferred from equation (\ref{Lx}) and thus
increase the GW frequency of the signal, shifting the curves to the
right. From the leading term in equation (\ref{Omega3}), it is clear
that the total GW power is simply proportional to $f_{\rm co}$, so
increasing this parameter linearly increases the total amplitude of the
GW spectrum.

\section{EVENT RATES AND RESOLVABLE SIGNALS}

The GW spectra calculated from equation (\ref{Omega3}) represent the {\it
time-averaged} signal from all the AGN in the universe, but at any
single time, there may only be a few sources that are emitting at a
given frequency. This can be seen from comparing Figures
\ref{single_inspiral} and \ref{fiducial}: while the signal-to-noise
ratio of a single inspiral can be very large around $1-10$ mHz, a
relatively small fraction of the inspiral time is spent in that band,
significantly reducing the time-averaged strain amplitude.

One way of estimating the number of individual signals at a given
frequency is by calculating the duty cycle $D(f)$:
\begin{equation}\label{D_f}
  D(f) = \frac{f_{\rm co}}{f_{\rm X} m \eta_{\rm gw}}
  \int dz \frac{dV}{dz}
  \int_{L_{\rm min}}^{L_{\rm max}} dL_{\rm X} \frac{dn}{d\ln L_{\rm X}}
  t_{\rm coh}(f_z),
\end{equation}
where the cosmological volume element is $dV/dz=4\pi r^2(z)/H(z)$ and
the ``coherence time'' $t_{\rm coh}(f)$ is approximated by the
Newtonian limit of the radiation reaction formula (e.g.\
\cite{peters}):
\begin{equation}
  t_{\rm coh}(f) \equiv \frac{f}{df/dt} \simeq \frac{5}{144}m^{-1} M^{-2/3}
  (\pi f)^{-8/3}.
\end{equation}

\begin{figure}
\caption{\label{duty_cycle} Duty cycle $D(f)$ for the nominal
  parameter values and a range of compact object masses $m$. When
  $D(f)\lesssim 1$, the inspiral signals should be individually
  resolvable.} 
\scalebox{0.6}{\includegraphics{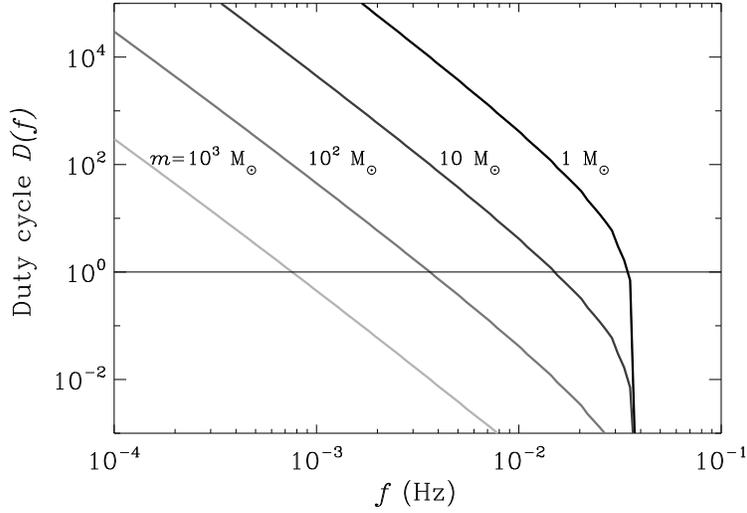}}
\end{figure} 

In Figure \ref{duty_cycle} we show the duty cycle for the fiducial
model parameters and a range of compact object masses $m$. The sharp
cutoff around 40 mHz is not physical, but rather due to the somewhat
artificial low-end
cutoff in the luminosity function, corresponding to a minimum value
for $M$ and thus maximum attainable frequency. Note that $D(f)$ is
proportional to $m^{-2}$, since smaller $m$ means more compact
objects, and also slower inspiral rates, thus each source spends more
time around a given $f$.

\begin{figure}
\caption{\label{unresolvable} The fiducial time-averaged signal ({\it top
  curve}), along with the unresolvable portions of the GW spectra, after
  subtracting out individual events with integrated SNR above a
  threshold of 15.}
\scalebox{0.6}{\includegraphics{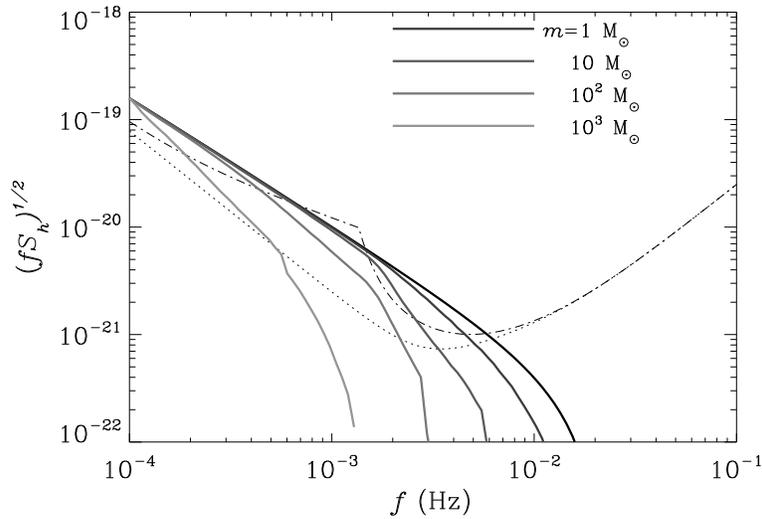}}
\end{figure} 

It is important to understand that $D(f)$ represents the total number
of sources in the observable universe, not all of which would be
individually resolvable with LISA. To estimate the subtractable portion of the
signal, we integrate the total signal-to-noise ratio (SNR) for a
three-year mission lifetime (see, e.g., \cite{finn}). For a given $m$, $M$, and source distance
$d$, we can calculate the minimum frequency above which the source
would be detectable with SNR above a certain threshold (here we use
15). Any contribution to equation (\ref{Omega3}) above that frequency
is ignored, resulting in an unresolvable GW spectrum,
plotted in Figure \ref{unresolvable}. As expected, with increasing
$m$, more inspiral events are detectable and the remaining
background confusion noise is diminished.


\begin{theacknowledgments}
  We thank Curt Cutler and Joe Silk for useful discussions.
\end{theacknowledgments}



\bibliographystyle{aipprocl} 


\begin{thebibliography}{9}

\bibitem{BH} P.L. Bender and D. Hils, Classical Quantum Gravity 
{\bf 14}, 1439 (1997).

\bibitem{Goodman:2003sf}
  J.~Goodman and J.~C.~Tan,
  Astrophys.\ J.\  {\bf 608}, 108 (2004).

\bibitem{Levin}
  Y.~Levin, Mon.\ Not.\ Royal\ Acad.\ Sci.\ submitted (2006)
  [arXiv:astro-ph/0603583].

\bibitem{sigl}
  G.~Sigl, J.~D.~Schnittman, and A.~Buonanno,
  Phys.\ Rev.\ D submitted (2006).

\bibitem{uaom}
  Y.~Ueda, M.~Akiyama, K.~Ohta, and T.~Miyaji,
  Astrophys.\ J.\  {\bf 598}, 886 (2003).

\bibitem{barrack}
  L.~Barrack and C.~Cutler,
  Phys.\ Rev.\ D {\bf 70}, 122002 (2004).

\bibitem{yu}
  Q.~j.~Yu and S.~Tremaine,
  Mon.\ Not.\ Roy.\ Astron.\ Soc.\  {\bf 335}, 965 (2002).

\bibitem{thorne}
  K.~S.~Thorne,
  Astrophys.\ J.\ {\bf 191}, 507 (1974).

\bibitem{marconi}
  A.~Marconi, G.~Risaliti, R.~Gilli, L.~K.~Hunt, R.~Maiolino and M.~Salvati,
  Mon.\ Not.\ Roy.\ Astron.\ Soc.\  {\bf 351}, 169 (2004).

\bibitem{ressell}
  M.~T.~Ressell and M.~S.~Turner, Comments Astrophys.\ {\bf 14}, 323
  (1990).
 
\bibitem{phinney}
  E.~S.~Phinney, [astro-ph/0108028].

\bibitem{finn}
  L.~S.~Finn and K.~S.~Thorne,
  Phys.\ Rev.\ D {\bf 62}, 124021 (2000).

\bibitem{peters} 
  P.\ C.\ Peters, Phys.\ Rev.\ {\bf 136}, B1224 (1964).

\end{thebibliography}


\end{document}